\documentclass[RNAAS]{aastex63}

\newcommand\omicron{o}

\received{20 January 2020}
\accepted{27 January 2020}
\submitjournal{RNAAS}

\shorttitle{A catalog of Galactic RSG+B systems}
\shortauthors{Pantaleoni Gonz\'alez \& Ma{\'\i}z Apell\'aniz}

\begin{document}

\title{A catalog of Galactic multiple systems with a red supergiant and a B star}

\correspondingauthor{J. Ma{\'\i}z Apell\'aniz}
\email{jmaiz@cab.inta-csic.es}

\author[0000-0001-9933-1229]{M. Pantaleoni Gonz\'alez}
\affiliation{Centro de Astrobiolog\'{\i}a, CSIC-INTA, E-28\,692 Villanueva de la Ca\~nada, Spain}
\affiliation{ Universidad Complutense de Madrid, E-28\,040 Madrid. Spain.}

\author[0000-0003-0825-3443]{J. Ma{\'\i}z Apell\'aniz}
\affiliation{Centro de Astrobiolog\'{\i}a, CSIC-INTA, E-28\,692 Villanueva de la Ca\~nada, Spain}

\author[0000-0003-1086-1579]{R. H. Barb\'a}
\affiliation{Universidad de La Serena, La Serena, Chile}

\author[0000-0003-1952-3680]{I. Negueruela}
\affiliation{Universidad de Alicante, E-03\,690 San Vicente del Raspeig, Spain}



%
%

\keywords{Binary stars --- Catalogs --- Early-type stars --- Late-type stars --- Supergiant stars}


\section{Introduction} \label{sec:intro}

Binary stars are useful to study many aspects of stellar structure and evolution and the systems where the two objects have very
different spectral types are especially so, as (barring interactions between them or with third objects) they signal coeval
evolutionary phases. In this context, \citet{Neugetal19}, from now on N19, indicate that ``until a year ago [2018], the total number of known Galactic 
binary RSGs was 11" (RSG = red supergiant) and then go on to wonder about where the missing binaries are and to point towards 87 new RSG+B binary systems in
M31, M33, the SMC, and the LMC. In this work we examine that quoted text.

\section{Methods} \label{sec:methods}

We have searched the literature and Simbad to find Galactic RSG+B binary systems. More specifically, we define RSGs as objects with spectral type G, K, or M
and luminosity classes II to Ia for consistency with N19, but we note that criterion includes some Cepheids, which can be of spectral type G-K0 in some phases. 
Results are listed in Table 1 along with some useful information. The separation column is populated 
if a visual component with $\Delta m$ small enough to contribute to the spectral type at short wavelengths is listed in the Washington Double Star Catalog 
\citep{Masoetal01}.

\section{Results} \label{sec:results}

We have found 108 Galactic RSG+B binary systems in the literature, with 61\% of the sample already present in two references alone \citep{GineCarq02,Samuetal17}. 
A significant number of the objects have two entries in the HD catalog, a likely consequence of the composite
nature of the spectra. The total number in the sample has to be taken with care, as literature spectral classifications can include gross 
errors (see e.g. \citealt{Maizetal16}). Also, RSGs in the spectral classification sense we are using here 
should not be identified with RSGs in the stellar evolution sense: some in our list are Cepheids and others may be
bright AGB stars or even more peculiar objects (e.g. $\beta$~Cyg~A, \citealt{BastAnto18}, or V838~Mon, \citealt{Munaetal07}). Nevertheless, the number of known Galactic
RSG+B systems is much larger than the one claimed by N19 and indeed it is not that different from the value found by those authors for the other four large galaxies in the Local Group. Nine of the 
eleven stars in N19 are in our sample. \object[AL Vel]{AL~Vel} and \object[Algol]{Algol} are missing because they do not satisfy the luminosity class requirement.

Obviously, there must be many more Galactic RSG+B systems to be discovered. Two biases are a sign of this. 71\% of the sample is brighter than $V = 8$ and 67\% is in the northern
hemisphere (despite the southern hemisphere containing many more Galactic disk stars). Future studies are required to find them and improve our statistics on these important systems.
Interestingly, there are no known Galactic RSG+O binary systems. There are, however, known Galactic BSG+O systems (e.g. \object[HD 115071]{HD~115\,071}, \citealt{Sotaetal14},
BSG = B-type supergiant) and at least one ASG+O system (\object[6 Cas]{6~Cas}, \citealt{Bartetal94}, ASG = A-type supergiant). This is a likely manifestation of the H-D limit 
\citep{HumpDavi79}.

\acknowledgments

(a) M.P.G. and J.M.A. and (b) I.N. acknowledge support from the Spanish Government Ministerio de Ciencia, Innovaci\'on y Universidades 
through grants (a) PGC2018-095\,049-B-C22 and (b) PGC2018-093\,741-B-C21, respectively. R.H.B. acknowledges support from DIDULS Project 18\,143 
and the ESAC visitors program. This research has made extensive use of the SIMBAD database, operated at CDS, Strasbourg, France.

\bibliography{general}{}
\bibliographystyle{aasjournal}



\begin{longrotatetable}
\begin{deluxetable*}{lllllllllrrrrll}
\setlength{\tabcolsep}{4pt}
\tablenum{1}
\tablecaption{Objects, spectral classifications, Galactic coordinates, $V$ magnitudes, separations, references, and comments, sorted by $l$.\label{tab:main}}
\tablewidth{0pt}
\tablehead{
\colhead{Name} & \colhead{Alternate name} & \multicolumn{3}{c}{RSG} && \multicolumn{3}{c}{B star} & \colhead{$l$}    & \colhead{$b$}    & \colhead{$V$}    & \colhead{sep.}      & \colhead{Ref.} & \colhead{Comments} \\
\cline{3-5}\cline{7-9}
\colhead{}     & \colhead{}               & ST & LC & Q.            && ST & LC & Q.               & \colhead{(deg.)} & \colhead{(deg.)} & \colhead{}       & \colhead{(\arcsec)} & \colhead{}     & \colhead{}         
}
\startdata
\object[HD 161387]{HD 161\,387}     & V777 Sgr                  & K5    & Ib     & \nodata && B6:      & \nodata & \nodata &   2.43 &     1.27 &     7.5 & \nodata & P98 &                                               \\ 
\object[HD 173297]{HD 173\,297}     & V350 Sgr                  & F5/G2 & Ib     & \nodata && B9       & V       & \nodata &  13.76 &  $-$7.96 &     7.5 & \nodata & S17 & Cepheid                                       \\ 
\object[HD 168701]{HD 168\,701}     & V4390 Sgr                 & K3    & II     & \nodata && B3/5?    & \nodata & \nodata &  15.08 &  $-$1.04 &     7.7 & \nodata & G02 &                                               \\
\object[FR Sct]{FR Sct}             & HIP 90\,115               & M2.5  & Iab    & ep      && B        & \nodata & \nodata &  18.47 &     0.34 &    11.6 & \nodata & S17 &                                               \\
\object[Dabih Major]{$\beta^1$ Cap} & Dabih Major               & G9    & II:    & \nodata && B8       & \nodata & p:Si:   &  29.15 & $-$26.37 &     3.1 &     0.1 & G02 & Visual and spectroscopic binary               \\
\object[HD 169689]{HD 169\,689}     & V2291 Oph                 & G9    & II     & \nodata && B9       & V       & \nodata &  37.21 &     9.36 &     5.7 & \nodata & G02 &                                               \\
\object[eta Aql]{$\eta$ Aql}        & 55 Aql                    & F6-G2 & Ib     & \nodata && B9.8     & V       & \nodata &  40.93 & $-$13.08 &     3.8 &     0.7 & E13 & Cepheid                                       \\ 
\object[chi Aql]{$\chi$ Aql}        & 47 Aql                    & G2    & Ib     & \nodata && B5.5     & \nodata & \nodata &  49.34 &  $-$5.70 &     5.3 &     0.4 & G02 &                                               \\
\object[delta Sge]{$\delta$ Sge}    & 7 Sge                     & M2.5  & II-III & \nodata && B9       & V       & \nodata &  55.77 &  $-$3.38 &     3.8 &     0.1 & S17 & Visual and spectroscopic binary               \\
\object[HD 187321]{HD 187\,321}     & BD $+$18 4252             & G5    & Ib-II: & \nodata && B7       & IV      & \nodata &  56.21 &  $-$3.49 &     7.1 &     0.4 & G02 &                                               \\ 
\object[HD 190361]{HD 190\,361}     & BD $+$20 4406             & K4    & Ib     & \nodata && B4       & IV-V    & \nodata &  59.93 &  $-$5.41 &     7.3 & \nodata & G02 &                                               \\ 
\object[HD 187299]{HD 187\,299}     & BD $+$24 3889             & G8    & Ib     & \nodata && B6/9     & V       & \nodata &  61.48 &  $-$0.31 &     7.3 & \nodata & S14 &                                               \\
\object[HD 199738]{HD 199\,378}     & BD $+$14 4478             & K1.5  & II     & \nodata && B7:      & \nodata & \nodata &  61.83 & $-$19.25 &     7.3 & \nodata & G02 &                                               \\
\object[beta Cyg A]{$\beta$ Cyg A}  & Albireo A                 & K3    & II     & \nodata && B9       & V       & \nodata &  62.11 &     4.57 &     3.1 &     0.3 & B18 & Red star has low mass?                        \\
\object[HD 186518]{HD 186\,518}     & PS Vul                    & K3    & II     & \nodata && B6       & \nodata & \nodata &  62.82 &     1.61 &     6.4 &     0.3 & G02 & Visual and eclipsing binary                   \\
\object[22 Vul]{22 Vul}             & HD 192\,713               & G4    & Ib     & \nodata && B7/9?    & \nodata & \nodata &  63.47 &  $-$6.36 &     5.2 & \nodata & G02 & In \citet{Neugetal19}                         \\
\object[BD +27 3542]{BD $+$27 3542} & Tyc 2148-00114-1          & K3    & Iab    & \nodata && B3       & II      & \nodata &  64.02 &     0.18 &     8.7 & \nodata & N20 &                                               \\ 
\object[HD 186688]{HD 186\,688}     & SU Cyg                    & F2-G0 & I-II   & \nodata && B7       & V       & \nodata &  64.76 &     2.51 &     6.4 & \nodata & S17 & Cepheid                                       \\
\object[HD 186097]{HD 186\,097}     & BD $+$30 3692             & G1:   & Ib-II  & \nodata && B8       & III     & \nodata &  65.65 &     3.90 &     7.3 &     0.8 & G02 &                                               \\ 
\object[HD 196753]{HD 196\,753}     & BD $+$23 4085             & K1    & II     & \nodata && B7       & \nodata & \nodata &  66.65 & $-$10.61 &     5.9 & \nodata & G02 &                                               \\
\object[V2028 Cyg]{V2028 Cyg}       & ALS 10\,651               & K2    & Ib/II  & \nodata && B4       & III     & [e]     &  67.63 &     1.26 &    10.9 & \nodata & Z01 &                                               \\
\object[V2417 Cyg]{V2417 Cyg}       & AS 381                    & K:    & I:     & \nodata && B1       & \nodata & [e]     &  70.58 &     0.57 &    14.4 & \nodata & M02 &                                               \\ 
\object[47 Cyg]{47 Cyg}             & HD 196\,093               & K6:   & Ib     & \nodata && B2.5:    & \nodata & \nodata &  75.43 &  $-$2.93 &     4.6 &     0.3 & G02 & Visual and spectroscopic binary               \\
\object[HD 193469]{HD 193\,469}     & BD $+$38 4003             & K4.5  & Ib     & \nodata && B8       & V       & \nodata &  76.76 &     1.68 &     6.3 &     3.5 & G02 &                                               \\ 
\object[31 Cyg]{$\omicron^1$ Cyg}   & 31 Cyg                    & K4    & Iab    & \nodata && B4       & IV-V    & \nodata &  82.68 &     6.78 &     3.8 & \nodata & S17 & In \citet{Neugetal19}                         \\
\object[32 Cyg]{$\omicron^2$ Cyg}   & 32 Cyg                    & K5    & Iab    & \nodata && B4       & IV-V    & \nodata &  83.67 &     7.05 &     4.0 & \nodata & S17 & In \citet{Neugetal19}                         \\
\object[V1068 Cyg]{V1068 Cyg}       & BD $+$41 4100             & G8    & II-III & \nodata && B9       & \nodata & \nodata &  86.69 &  $-$5.28 &    10.3 & \nodata & S17 &                                               \\
\object[HD 205114]{HD 205\,114}     & BD $+$51 3079             & G2    & Ib     & \nodata && B7/8     & IV:     & \nodata &  95.22 &     0.86 &     6.2 & \nodata & G02 &                                               \\
\object[HD 203338]{HD 203\,338}     & V381 Cep                  & M1    & Ib     & ep      && B2       & \nodata & ep      &  98.18 &     6.36 &     5.8 &     0.1 & M69 &                                               \\
\object[5 Lac]{5 Lac}               & HD 213\,310               & K6-M0 & I      & \nodata && B7/8?    & IV:     & \nodata &  99.66 &  $-$8.65 &     4.4 & \nodata & G02 &                                               \\
\object[HDE 235749]{HDE 235\,749}   & BD $+$54 2698             & M2    & Ib     & \nodata && B        & \nodata & \nodata & 101.47 &  $-$0.80 &     8.9 & \nodata & D18 &                                               \\ 
\object[ALS 12387]{ALS 12\,387}     & LS III $+$57 47           & K3    & Ib     & \nodata && B        & \nodata & \nodata & 104.52 &     0.75 &    10.7 &     6.3 & D18 & {\it Gaia} DR2 vis. comp. not in WDS          \\ 
\object[HD 208816]{HD 208\,816}     & VV Cep                    & M2    & Ia-Iab & ep      && B8:      & V       & e       & 104.92 &     7.05 &     4.9 & \nodata & S17 & In \citet{Neugetal19}                         \\
\object[delta Cep]{$\delta$ Cep}    & 27 Cep                    & F5-G2 & Ib     & \nodata && B7-8     & \nodata & \nodata & 105.19 &     0.53 &     3.8 & \nodata & E13 & Prototype cepheid                             \\ 
\object[U Lac]{U Lac}               & HD 215\,924               & M4    & Iab    & ep      && B        & \nodata & \nodata & 105.82 &  $-$3.55 &     9.4 & \nodata & S17 &                                               \\
\object[HD 214369]{HD 214\,369}     & W Cep                     & K0-M2 & Ia     & ep      && B0/B1    & \nodata & \nodata & 106.02 &     0.06 &     7.6 & \nodata & S17 &                                               \\
\object[HD 218393]{HD 218\,393}     & BD $+$49 4045             & G8    & II     & \nodata && B3       & \nodata & pe      & 106.36 &  $-$9.29 &     6.9 & \nodata & P04 &                                               \\ 
\object[HD 217476]{HD 217\,476}     & V509 Cas                  & F8-K  & Ia-0   & e       && B1       & V       & \nodata & 108.16 &  $-$2.70 &     5.1 & \nodata & S17 &                                               \\
\object[HD 213503]{HD 213\,503}     & BD $+$67 1443             & K2    & II:    & \nodata && B8       & IV:     & \nodata & 110.36 &     8.88 &     7.9 & \nodata & G02 &                                               \\
\object[psi And]{$\psi$ And}        & 20 And                    & G5    & Ib     & \nodata && B9/A0    & IV:     & \nodata & 111.34 & $-$14.97 &     5.0 &     0.3 & G02 &                                               \\
\object[KN Cas]{KN Cas}             & BD $+$61 8                & M1    & Ib     & ep      && B3       & V       & \nodata & 118.15 &     0.19 &     9.5 &     0.2 & S17 &                                               \\
\object[V641 Cas]{V641 Cas}         & BD $+$63 3                & M3    & Iab    & e       && B2.5     & \nodata & \nodata & 118.34 &     1.46 &     8.3 & \nodata & S17 &                                               \\
\object[HDE 236429]{HDE 236\,429}   & DL Cas                    & F5-G2 & Ib     & \nodata && B9       & V       & \nodata & 120.27 &  $-$2.55 &     8.6 & \nodata & S17 & Cepheid                                       \\
\object[HD 3210]{HD 3210}           & BD $+$48 177              & K4    & II-III & \nodata && B2       & \nodata & \nodata & 120.30 &  $-$6.63 &     7.0 &     0.3 & G02 & Visual and spectroscopic binary               \\ 
\object[V554 Cas]{V554 Cas}         & BD $+$61 219              & M2    & I      & \nodata && B        & \nodata & e       & 125.11 &  $-$0.28 &     9.5 & \nodata & S17 &                                               \\
\object[BD +59 224]{BD $+$59 224}   & Tyc 4030-00149-1          & K4.5  & Ib     & \nodata && B3       & V       & \nodata & 126.13 &  $-$2.33 &     9.5 & \nodata & G04 & In \citet{Neugetal19}                         \\
\object[HD 9352]{HD 9352}           & BD $+$57 320              & K3    & Ib-II  & \nodata && B7/8:    & \nodata & \nodata & 128.44 &  $-$4.09 &     5.7 & \nodata & G02 &                                               \\
\object[AZ Cas]{AZ Cas}             & BD $+$60 310              & M0    & Ib     & e       && B0-B1    & V       & \nodata & 128.97 &  $-$0.85 &     9.2 & \nodata & S17 &                                               \\
\object[55 Cas]{55 Cas}             & HD 13\,474                & G0    & II-III & \nodata && B9       & V       & \nodata & 131.08 &     4.98 &     6.1 &     0.1 & M69 &                                               \\
\object[HD 12401]{HD 12\,401}       & XX Per                    & M4    & Ib     & \nodata && B        & \nodata & \nodata & 133.10 &  $-$6.22 &     8.2 & \nodata & S17 & In \citet{Neugetal19}                         \\
\object[gamma And]{$\gamma$ And}    & Almach                    & K3    & II     & \nodata && B9.5     & V       & \nodata & 136.96 & $-$18.56 &     2.1 &     9.4 & A95 & Triple system (next line)                     \\
                                    &                           &       &        &         && A0       & V       & \nodata &        &          &         &     9.4 & A95 &                                               \\
\object[HDE 237006]{HDE 237\,006}   & BD $+$57 641              & M1    & Ib     & e       && B:       & \nodata & \nodata & 138.02 &  $-$1.37 &     9.3 & \nodata & H69 &                                               \\
\object[HD 16082]{HD 16\,082}       & BD $+$51 599              & K0    & II     & \nodata && B6       & \nodata & \nodata & 138.94 &  $-$7.60 &     7.3 & \nodata & G02 &                                               \\
\object[HD 17306]{HD 17\,306}       & BD $+$53 574              & K3    & Iab    & \nodata && B:       & \nodata & \nodata & 139.65 &  $-$4.85 &     7.9 & \nodata & B57 &                                               \\
\object[HD 19278]{HD 19\,278}       & BD $+$56 779              & K2    & II     & \nodata && B7/8     & \nodata & \nodata & 140.97 &  $-$1.34 &     8.2 &     1.7 & G02 &                                               \\ 
\object[HD 23089]{HD 23\,089}       & BD $+$62 604              & G2    & Ib/II  & \nodata && B7       & III/IV  & \nodata & 141.10 &     6.81 &     4.8 & \nodata & G06 &                                               \\
\object[HD 24480]{HD 24\,480}       & BD $+$60 768              & K3    & II     & \nodata && B8/9     & \nodata & \nodata & 143.54 &     5.90 &     5.1 &     1.7 & G02 &                                               \\ 
\object[HD 17245]{HD 17\,245}       & BD $+$43 576              & G8    & II-III & \nodata && B8:      & \nodata & \nodata & 143.89 & $-$13.82 &     6.5 & \nodata & G02 &                                               \\
\object[HDE 237190]{HDE 237\,190}   & RW Cam                    & F5-G1 & Ib     & \nodata && B8       & III     & \nodata & 144.85 &     3.80 &     8.7 &     0.3 & S17 & Cepheid                                       \\
\object[HD 21771]{HD 21\,771}       & BD $+$44 732              & K3    & II     & \nodata && B8       & \nodata & \nodata & 150.56 &  $-$9.23 &     7.3 & \nodata & G02 &                                               \\
\object[HD 27395]{HD 27\,395}       & BD $+$49 1165             & G9    & II     & \nodata && B        & \nodata & \nodata & 153.60 &     0.12 &     7.2 &     2.0 & C05 & Visual and spectroscopic binary               \\
\object[mu Per]{$\mu$ Per}          & 51 Per                    & G0    & Ib     & \nodata && B9.5     & \nodata & \nodata & 153.94 &  $-$1.82 &     4.2 & \nodata & G02 &                                               \\ 
\object[f Per]{f Per}               & 52 Per                    & G9    & II:    & \nodata && B9.5/A0: & IV:     & \nodata & 159.45 &  $-$7.54 &     4.7 & \nodata & G02 &                                               \\
\object[BD +43 1401]{BD $+$43 1041} & ALS 8026                  & G5/K0 & II/III & \nodata && B8:      & V       & \nodata & 160.29 &  $-$1.48 &     8.7 & \nodata & S14 &                                               \\
\object[58 Per]{58 Per}             & HD 29\,094                & G7    & Ib     & \nodata && B8/9.5:  & \nodata & \nodata & 161.76 &  $-$4.03 &     4.3 & \nodata & G02 &                                               \\
\object[zeta Aur]{$\zeta$ Aur}      & 8 Aur                     & K5    & Ib-II  & \nodata && B6.5     & IV-V    & \nodata & 165.02 &  $-$0.43 &     3.8 & \nodata & S17 & In \citet{Neugetal19}                         \\
\object[HD 36947]{HD 36\,947}       & BD $+$43 1315             & G7    & Ib:    & \nodata && B7       & III:    & \nodata & 166.22 &     6.54 &     7.3 &     0.1 & G02 &                                               \\
\object[HD 33203]{HD 33\,203}       & BD $+$37 1067             & K4    & II:    & \nodata && B2       & II:     & \nodata & 168.95 &  $-$1.49 &     6.1 &     1.6 & P98 &                                               \\ 
\object[36 Tau]{36 Tau}             & HD 25\,555                & K1    & II     & \nodata && B7.5     & IV:     & \nodata & 169.67 & $-$20.79 &     5.5 &  $<$0.1 & G02 & Visual and spectroscopic binary               \\
\object[HD 27639]{HD 27\,639}       & BD $+$20 744              & K5:   & II     & \nodata && B7/8     & \nodata & \nodata & 175.26 & $-$19.98 &     6.0 &     1.9 & G02 &                                               \\ 
\object[HDE 246901]{HDE 246\,901}   & BD $+$33 1138             & G5:   & Ib:    & \nodata && B1:      & \nodata & \nodata & 176.00 &     2.26 &     8.1 & \nodata & M55 &                                               \\
\object[HD 42474]{HD 42\,474}       & WY Gem                    & M2    & Iab    & ep      && B2-B3    & V       & \nodata & 187.91 &     2.26 &     7.4 & \nodata & S17 &                                               \\
\object[HD 39286]{HD 39\,286}       & BD $+$19 1110             & K0    & IIb    & \nodata && B8       & IV      & \nodata & 188.57 &  $-$3.34 &     6.0 & \nodata & G00 &                                               \\
\object[HD 47086]{HD 47\,086}       & BD $+$23 1433             & G:    & I      & \nodata && B/A      & \nodata & \nodata & 190.31 &     7.72 &     6.7 &     0.1 & B83 &                                               \\
\object[HD 44990]{HD 44\,990}       & T Mon                     & G8    & Iab-Ib & \nodata && B9.8     & V       & \nodata & 203.63 &  $-$2.56 &     6.0 & \nodata & E94 & Cepheid                                       \\ 
\object[HD 39118]{HD 39\,118}       & BD $+$01 1148             & K0    & II     & \nodata && B7/8     & \nodata & \nodata & 204.02 & $-$12.57 &     6.0 & \nodata & G02 &                                               \\
\object[HD 45910]{HD 45\,910}       & AX Mon                    & K1    & II-III & \nodata && B1       & IV      & eq      & 205.33 &  $-$1.95 &     6.9 & \nodata & S17 &                                               \\
\object[HD 50820]{HD 50\,820}       & BD $-$01 1446             & K2    & II     & \nodata && B3       & IV      & e       & 214.87 &  $-$0.08 &     6.2 & \nodata & H82 &                                               \\
\object[HD 52690]{HD 52\,690}       & V926 Mon                  & M2    & Ib     & \nodata && B7/9?    & IV:     & \nodata & 217.49 &     0.66 &     6.6 & \nodata & G02 &                                               \\
\object[V838 Mon]{V838 Mon}         & Nova Mon 2002             & M     & I      & \nodata && B3       & V       & \nodata & 217.80 &     1.05 &     var & \nodata & M07 & In \citet{Neugetal19}                         \\
\object[HD 55684]{HD 55\,684}       & BD $-$04 1862             & K3    & II     & \nodata && B7.5     & III:    & \nodata & 220.07 &     2.59 &     7.3 & \nodata & G02 &                                               \\
\object[mu CMa]{$\mu$ CMa}          & 18 CMa                    & K3    & II     & \nodata && B8.5     & \nodata & \nodata & 225.99 &  $-$5.34 &     4.9 &     2.8 & G02 &                                               \\
\object[BD -12 1805]{BD $-$12 1805} & NGC 2345-34               & K3:   & II     & \nodata && B        & \nodata & \nodata & 226.57 &  $-$2.29 &     9.9 & \nodata & M74 &                                               \\
\object[HD 59067]{HD 59\,067}       & BD $-$11 1951             & G4:   & Ib     & \nodata && B2       & \nodata & \nodata & 227.37 &     2.67 &     5.9 &     0.7 & G02 &                                               \\ 
\object[HD 60415]{HD 60\,415}       & KQ Pup                    & M2    & Iab    & ep      && B2       & V       & \nodata & 230.67 &     2.52 &     5.0 & \nodata & S17 &                                               \\
\object[F Hya]{F Hya}               & HD 74\,395                & G1    & Ib     & \nodata && B9.5/A0  & \nodata & \nodata & 233.29 &    20.97 &     4.6 & \nodata & G02 &                                               \\
\object[pi Pup]{$\pi$ Pup}          & HD 56\,855                & K3    & Ib     & \nodata && B5       & \nodata & \nodata & 249.01 & $-$11.28 &     2.7 & \nodata & S14 &                                               \\
\object[V624 Pup]{V624 Pup}         & CD $-$32 4694             & M2    & Iab    & \nodata && B1       & V       & \nodata & 249.58 &  $-$1.42 &    10.9 & \nodata & S17 &                                               \\
\object[HD 31244]{HD 31\,244}       & CPD $-$51 600             & K3    & II-III & \nodata && B5       & \nodata & \nodata & 259.03 & $-$39.32 &     6.6 & \nodata & J60 &                                               \\
\object[HD 81137]{HD 81\,137}       & WY Vel                    & M3    & Ib:    & ep      && B        & \nodata & \nodata & 274.14 &  $-$1.82 &     8.8 & \nodata & S17 &                                               \\
\object[NGC 3105 24]{NGC 3105-24}   & [AMN2018] 348             & K3    & Ib     & \nodata && B2       & V:      & \nodata & 279.91 &     0.28 &    13.0 & \nodata & A18 &                                               \\ 
\object[HDE 300933]{HDE 300\,933}   & CPD $-$56 3586            & M2    & Iab/Ib & \nodata && B        & \nodata & \nodata & 285.42 &     1.45 &     8.3 & \nodata & H72 &                                               \\
\object[HDE 303344]{HDE 303\,344}   & CPD $-$57 3805            & M2    & Ib     & \nodata && B        & \nodata & \nodata & 287.08 &     1.09 &     9.3 & \nodata & D79 &                                               \\
\object[HD 93281]{HD 93\,281}       & V730 Car                  & M1    & Iab    & \nodata && B        & \nodata & \nodata & 287.70 &  $-$0.86 &     7.8 & \nodata & S17 &                                               \\
\object[HD 101007]{HD 101\,007}     & CPD $-$60 3178            & M3    & Ib     & \nodata && B        & \nodata & \nodata & 294.08 &     0.40 &     7.0 & \nodata & K89 &                                               \\
\object[HD 101947]{HD 101\,947}     & V810 Cen                  & F5-G0 & Ia-0   & \nodata && B1       & Iab     & \nodata & 295.18 &  $-$0.64 &     5.0 & \nodata & S17 &                                               \\
\object[HD 101712]{HD 101\,712}     & V772 Cen                  & M2    & Ib     & ep      && B        & \nodata & \nodata & 295.24 &  $-$1.59 &     7.9 & \nodata & S17 &                                               \\
\object[CD -61 3575]{CD $-$61 3575} & Tyc 8992-00314-1          & M2    & Ia     & ep      && B        & \nodata & \nodata & 302.09 &     0.92 &     9.9 & \nodata & S14 &                                               \\
\object[KN Cen]{KN Cen}             & HIP 66\,383               & G8    & Iab    & \nodata && B6       & V       & \nodata & 307.76 &  $-$2.11 &     9.9 &     0.3 & S17 & Cepheid                                       \\
\object[HD 119756]{HD 119\,796}     & V766 Cen                  & G8    & Ia-0   & \nodata && B0       & Ib      & p:      & 309.30 &  $-$0.41 &     6.8 &     0.1 & S17 & In \citet{Neugetal19}                         \\
\object[26 Cir]{26 Cir}             & HD 130\,702               & F2-G2 & II     & \nodata && B4       & \nodata & \nodata & 315.83 &  $-$4.01 &     6.0 &     0.3 & S17 & Cepheid                                       \\
\object[HD 134270]{HD 134\,270}     & CPD $-$54 6367            & G2/5  & Ib     & \nodata && B8       & V       & \nodata & 321.96 &     2.27 &     5.4 & \nodata & N85 &                                               \\
\object[ALS 3371]{CPD $-$58 6053}   & ALS 3371                  & M2    & Ia     & \nodata && B        & \nodata & \nodata & 322.79 &  $-$2.28 &     9.9 & \nodata & D79 &                                               \\
\object[HD 146323]{HD 146\,323}     & S Nor                     & F8/G0 & Ib     & \nodata && B9.5     & V       & \nodata & 327.75 &  $-$5.40 &     6.5 & \nodata & E13 & Cepheid                                       \\ 
\object[HD 135345]{HD 135\,345}     & CPD $-$41 7104            & G5    & Ia     & \nodata && B        & \nodata & \nodata & 329.98 &    13.66 &     5.2 &     0.2 & J60 &                                               \\
\object[HD 145415]{HD 145\,415}     & CPD $-$53 7442            & K2    & Ib     & \nodata && B        & \nodata & \nodata & 329.85 &  $-$2.18 &     8.9 & \nodata & A17 &                                               \\
\object[alpha Sco]{$\alpha$ Sco}    & Antares                   & M0.5  & Iab    & \nodata && B3       & V:      & \nodata & 351.95 &    15.06 &     0.9 &     3.2 & C84 &                                               \\ 
\object[HD 172991]{HD 172\,991}     & CPD $-$39 8163            & K3    & II     & \nodata && B7       & \nodata & \nodata & 356.00 & $-$15.77 &     5.4 & \nodata & J60 &                                               \\
\hline
\multicolumn{15}{l}{\begin{minipage}[t]{22 cm}\textbf{References:} A95: \citet{AbtMorr95}. A17: \citet{Alonetal17b}. A18: \citet{Alonetal18}. B57: \citet{Bide57}. B83: \citet{Bide83}. 
B18: \citet{BastAnto18}. C05: \citet{Carqetal05}. C84: \citet{Corb84}. D79: \citet{Dril79}. D18: \citet{Dordetal18}. E94: \citet{EvanLyon94}.
E13: \citet{Evanetal13}. G00: \citet{GrifGrif00}. G02: \citet{GineCarq02}. G04: \citet{GraySkif04}. G06: \citet{Grifetal06}.
H69: \citet{Hump69}. H72: \citet{Humpetal72}. 
H82: \citet{Hend82}. 
J60: \citet{JascJasc60}. K89: \citet{KeenMcNe89}. M55: \citet{Morgetal55}. M69: \citet{Mark69}. M74: \citet{Moff74}.
M02: \citet{Miroetal02}. M07: \citet{Munaetal07}. N85: \citet{Nooretal85}. N20: Negueruela et al. (2020, submitted).
P98: \citet{ParsAke98}. P04: \citet{Pouretal04}. S14: \citet{Skif14}. S17: \citet{Samuetal17}. Z01: \citet{Zick01}. \end{minipage}} \\
\enddata
\end{deluxetable*}
\end{longrotatetable}

\end{document}